\documentclass{ceurart}

\usepackage{listings}
\usepackage{graphicx}
\usepackage{comment}
\usepackage{hyperref}
\usepackage{url}
\usepackage{cleveref}
\usepackage{tcolorbox}
\usepackage{listings}
\usepackage{pifont}
\usepackage{makecell}
\usepackage{enumitem}
\usepackage{xspace}

\newcommand{\cmark}{\ding{51}}
\newcommand{\xmark}{\ding{55}}

\newcommand{\falconmodel}{\texttt{falcon\--3\--10b\--instruct}\xspace}
\newcommand{\phimodel}{\texttt{phi\--4}\xspace}
\newcommand{\qwenmodel}{\texttt{qwen\--2.5\--14b\--instruct}\xspace}
\newcommand{\gptomodel}{\texttt{gpt\--4o}\xspace}
\newcommand{\gptominimodel}{\texttt{gpt\--4o\--mini}\xspace}
\newcommand{\gptturbomodel}{\texttt{gpt\--4\--turbo}\xspace}
\newcommand{\gptpreviewmodel}{\texttt{gpt\--4.5\--preview\--2025\--02\--27}\xspace}

\definecolor{blue-frame}{RGB}{108,142,191}
\definecolor{blue-background}{RGB}{218,232,252}

\definecolor{codegreen}{rgb}{0,0.6,0}
\definecolor{codepurple}{rgb}{0.58,0,0.82}

\definecolor{codeborder}{rgb}{0.4, 0.4, 0.4}
\definecolor{codebackground}{rgb}{0.9608, 0.9608, 0.9608}
\definecolor{codefontcolor}{rgb}{0.2, 0.2, 0.2}

\lstdefinestyle{codestyle}{
    backgroundcolor=\color{codebackground},
    commentstyle=\color{codepurple},
    frame=leftline,
    breakatwhitespace=false,         
    breaklines=true,                 
    captionpos=t,                    
    keepspaces=true,
    showstringspaces=false,
    numbersep=5pt,             
    tabsize=2,
    basicstyle=\ttfamily\scriptsize,
    rulecolor=\color{codeborder}
}
\begin{document}

\copyrightyear{2025}
\copyrightclause{Copyright for this paper by its authors. Use permitted under Creative Commons License Attribution 4.0 International (CC BY 4.0).}
\conference{HAIC 2025 : First International Workshop on Human-AI Collaborative Systems, October 25 -- 30, Bologna, Italy}

\title{Comparative Analysis of Large Language Models for the Machine-Assisted Resolution of User Intentions}

\author[1]{Justus Flerlage}[orcid=0009-0007-2929-3408, email=j.flerlage@tu-berlin.de]
\cormark[1]
\author[2]{Alexander Acker}[orcid=0000-0002-0108-3034, email=alexander.acker@logsight.ai]
\author[1]{Odej Kao}[orcid=0000-0001-6454-6799, email=odej.kao@tu-berlin.de]

\address[1]{Distributed and Operating Systems Group\\Technische Universität Berlin\\Berlin, Germany}
\address[2]{logsight.ai GmbH\\Berlin, Germany}

\cortext[1]{Corresponding author.}

\begin{abstract}
Large Language Models (LLMs) have emerged as transformative tools for natural language understanding and user intent resolution, enabling tasks such as translation, summarization, and, increasingly, the orchestration of complex workflows. This development signifies a paradigm shift from conventional, GUI-driven user interfaces toward intuitive, language-first interaction paradigms. Rather than manually navigating applications, users can articulate their objectives in natural language, enabling LLMs to orchestrate actions across multiple applications in a dynamic and contextual manner. However, extant implementations frequently rely on cloud-based proprietary models, which introduce limitations in terms of privacy, autonomy, and scalability. For language-first interaction to become a truly robust and trusted interface paradigm, local deployment is not merely a convenience; it is an imperative.    This limitation underscores the importance of evaluating the feasibility of locally deployable, open-source, and open-access LLMs as foundational components for future intent-based operating systems. In this study, we examine the capabilities of several open-source and open-access models in facilitating user intention resolution through machine assistance. A comparative analysis is conducted against OpenAI's proprietary GPT-4-based systems to assess performance in generating workflows for various user intentions. The present study offers empirical insights into the practical viability, performance trade-offs, and potential of open LLMs as autonomous, locally operable components in next-generation operating systems. The results of this study inform the broader discussion on the decentralization and democratization of AI infrastructure and point toward a future where user-device interaction becomes more seamless, adaptive, and privacy-conscious through locally embedded intelligence.
\end{abstract}
\begin{keywords}
    User-Machine Interaction\sep Large Language Models\sep Artificial Intelligence\sep Code Generation\sep GUI-less Operating Systems
\end{keywords}
\maketitle
\section{Introduction}
Contemporary LLMs possess the capability to comprehend natural language, discern user intent from input expressions, and execute tasks such as document summarization or translation \cite{jin2024comprehensive}, image generation \cite{zhang2023controllable} or code generation \cite{zhang2023planninglargelanguagemodels} tasks. Beyond these functions, LLMs also present the potential to deconstruct complex intents into discrete, actionable steps, thereby enabling the automated construction of workflows in a manner analogous to human reasoning \cite{xie2024human}. LLMs have the potential to profoundly transform human-device interaction by supplanting rigid graphical interfaces with intuitive, conversational ones. Rather than navigating through menus or memorizing application-specific commands, users can articulate their objectives in natural language. LLMs are responsible for interpreting these inputs and orchestrating actions across various applications and services in a dynamic manner. As a consequence, complex tasks are simplified, and the system adapts to each user's habits and context, thereby personalizing the user experience. This shift is not only particularly salient in the context of mobile devices, where screen space and input methods are constrained, but in other applications for human-computer interaction as well, such as robotics, where robots mimic human-like communication. Interfaces are undergoing a paradigm shift towards invisible, language-first systems, whereby interaction resembles conversing with a smart assistant more than utilising a conventional device.
\\\\
For instance, current systems necessitate the manual coordination of multiple applications to reschedule an appointment. Despite the ostensible simplicity of the user-given intention "Reschedule my appointment for tonight," the process introduces a cumbersome and complicated workflow consisting of multiple steps. The user is required to manually open the calendar application and search for the appointment, locating the participants. The user is then prompted to access the contacts application to research the contact details of the relevant participants. This approach is adopted to facilitate effective communication via telephone or text message for the purpose of negotiating alternative dates. The process under discussion is characterized by its cumbersome and time-consuming nature, especially when incorporating multiple participants. Additionally, the user is required to devise a sequence of actions to operate the various applications, necessitating not only a fundamental understanding of the provided interfaces, but also the capacity to operate them successfully.
\\\\
The prevailing design of operating systems has been predicated for the aforementioned interaction mechanisms with GUIs, hierarchical file management, and the shell, allocating the responsibility for interaction to the user. Therefore, the interaction mechanisms initiated by LLMs necessitate a reconceptualization of fundamental design decisions in contemporary operating systems. In our previous work, we presented the first step on the path to such a GUI-less operating system with the utilization of the proprietary \gptominimodel model \cite{flerlage2025machinegeneratedcoderesolutionuser}. Nevertheless, the proposed methodology engenders a considerable degree of interdependence. The advent of future mobile devices is poised to achieve user intentions independently of external infrastructure. It is imperative to incorporate LLMs into local devices to ensure autonomy, privacy, extensibility, and optimization. Open-source and open-access models hold considerable potential in this endeavor and can serve as pivotal elements, not only for the future integration of such LLMs on local devices, but for the development of future operating systems with open and transparent ecosystems. This leads to the research question that guides this study, which is as follows: "How effective are open-source and open-access models in resolving user intentions for future intent-based operating systems, and what areas of research and development are indicated to enable broad, multi-domain deployment?".
\\\\
In this study, a comparative analysis of leading open-source and open-access models for this particular application domain is undertaken. We evaluate and analyze the performance of different LLMs for the purpose of generating different workflows for realizing a set of given user intentions. The comparison will include leading open-source and open-access models, such as Falcon 3, Phi 4, and Qwen 2, as well as proprietary models based on the fourth GPT generation from OpenAI, for comparison. We contribute our evaluation and analysis of the aforementioned LLMs, providing valuable insights regarding the feasibility of utilizing self-hosted, open-source, and open-access LLMs, as well as their comparative performance with proprietary models from OpenAI. The code for the experiments can be found in the Git repository at GitHub \footnote{https://github.com/dos-group/LLMWorkflowGenerator}.
\\\\
In the following section, \Cref{section:methodologies-and-approach}, we present the methodologies and the approach of our study. This is followed by  \Cref{section:experiments-and-results}, which the experiments and the results are presented. \Cref{section:discussion-and-result-interpretation} shows a discussion an interpretation of the results. The related work is shown afterwards in \Cref{section:related-work}. Finally, the conclusion is presented in \Cref{section:conclusion}.
\section{Methodologies and Approach}
\label{section:methodologies-and-approach}
\begin{figure}
    \centering
    \includegraphics[width=0.75\linewidth]{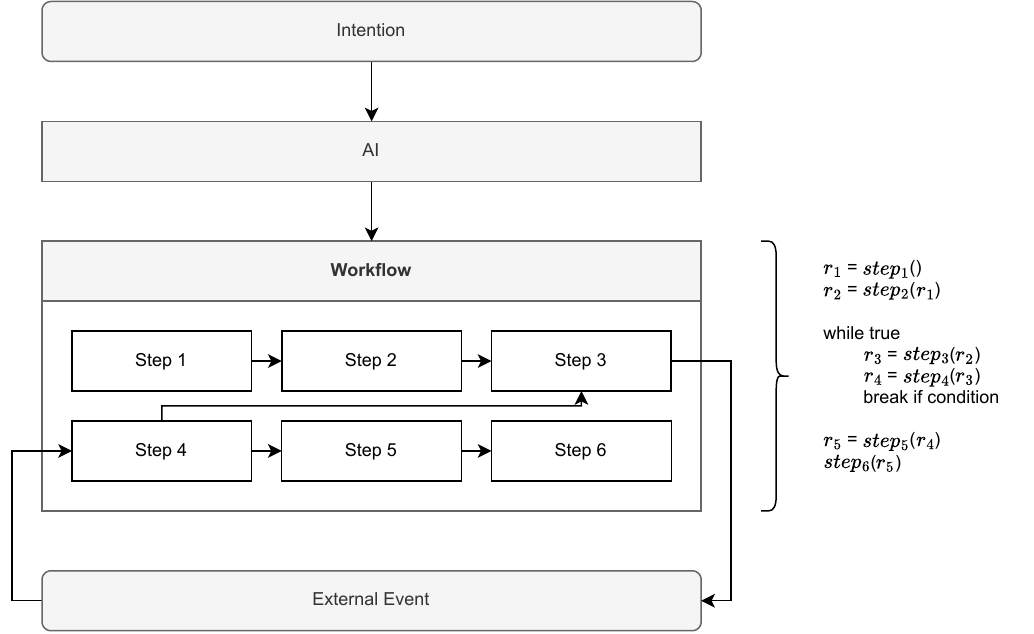}
    \caption{Workflows and state machines are analogous, and thus, they can be modeled using imperative programming languages.}
    \label{figure:workflow}
\end{figure}

\begin{figure}
    \centering
    \includegraphics[width=0.66\linewidth]{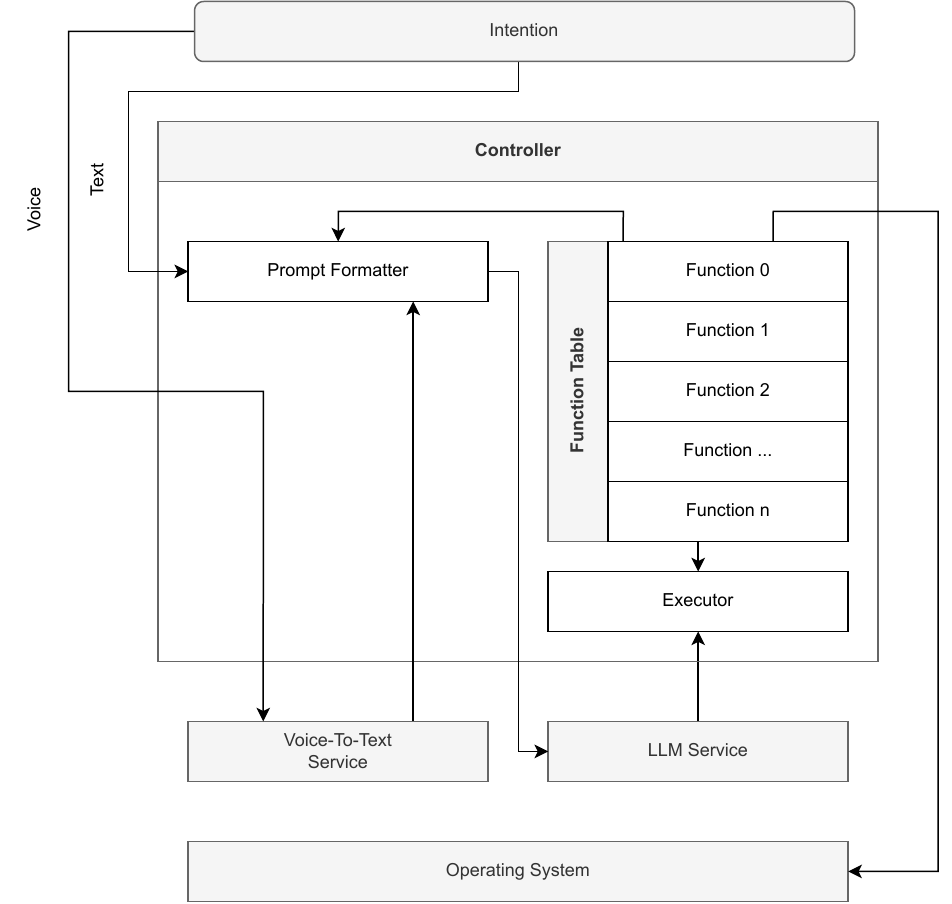}
    \caption{The deployed system architecture for our experiments.}
    \label{figure:system-architecture}
\end{figure}

The process of translating user intentions into actionable and executable workflows is of paramount importance for the development of future systems that prioritize intent-driven interaction mechanisms. Current LLMs have demonstrated the capacity to decompose user intentions into actionable steps, thereby enabling the design of workflows analogous to those employed by human users. The necessity of an intermediate representation is imperative for the description and modeling of these workflows and its necessary steps for resolving a given intention. This representation must possess the capacity to address arbitrary and complex user intentions.
\\\\
The code generation capabilities of LLMs are leveraged to synthesize workflows tailored to specific user intentions. These workflows are conceptualized as deterministic state machines, that can be effectively modeled using imperative programming languages, as shown in \Cref{figure:workflow}. The execution of such imperative programming language code is equivalent to state transitions of the state machine, which models the workflow. This refers to the ability to model both sequential steps and more complex control flow structures, such as loops and branches. Furthermore, it facilitates the interruption and preemption of steps and the management of asynchronous tasks, thereby enabling more flexible and dynamic program execution, and ultimately allowing for the incorporation of more complex user intentions. Within this framework, the LLM must not only interpret the user's high-level intent but also accurately comprehend and represent the underlying functionalities of the relevant application programming interface (API). This necessitates that the model parses the prompt with precision, analyzes the structure and semantics of the API, and subsequently generates syntactically correct and functionally coherent code that aligns with the intended behavior.
\\\\
\Cref{figure:system-architecture} provides a concise synopsis of the system architecture, which is the framework under discussion. It consists of an operating system running a dedicated \emph{Controller} application, which serves as the central coordination unit. The Controller is responsible for managing communication with an externally hosted LLM, acting as the interface between the local execution environment and the remote machine, which provides access to an LLM. In addition to overseeing interactions with LLM, the Controller orchestrates the scheduling, instantiation, and execution of workflows generated by the model. Subsequent to the synthesis of these workflows, they are integrated into the system's runtime environment in a seamless manner. This integration facilitates dynamic and adaptive responses to user intentions. The Controller's foundational element is the \emph{Function Table}, which contains a catalog of available functions, accompanied by their precise specifications, including signatures and associated implementation callbacks. It plays a crucial role in the generation of documentation, which is essential for guiding the LLM in generating valid and executable code. The Function Table, in conjunction with a textual representation of the user's intention, is employed by the \emph{Prompt Formatter} component to generate a prompt. This prompt is subsequently transmitted with a request to the \emph{LLM Service}. The system processes the user's prompt and generates the corresponding code, contingent on the user's intention and the available functions provided by the Function Table. Next, the \emph{Executor} employs the code generated by the LLM, subsequently executing it within a meticulously controlled environment. The execution of these functions is contingent upon the availability of the corresponding function implementations, which are stored in the Function Table. This measure is designed to prevent any unauthorized or unintended code execution and to establish an environment for executing the generated code in a controlled and isolated execution scope.
\\\\
The collection of metrics is conducted in accordance with the experimental protocol, encompassing the Time to First Token, the Response Time, as well as the inclusion of preambles, postambles, and code comments for measuring, comparing, and objectively evaluating the model's performance and responsiveness. The Time to First Token metric is defined as the duration required to receive the initial output from the specified LLM. This figure illustrates the model's initialization and processing overhead prior to generation. On the other hand, the Response Time metric is defined as the total time required to receive the complete output, which should be measured in seconds and occur within a few seconds to avoid disrupting the user's thought process \cite{nielsen1994usability}. Code elements, including comments, offer valuable insights into the decision-making process. However, these elements do so by increasing the response size, albeit to an infinitesimal degree. Preambles and postambles are integrated into the response and envelop the generated code block. These consist of explanations or introductory words from the LLM. The aforementioned elements are considered superfluous and serve only to augment the response size, thereby demonstrating an understanding of the designated role.
\\

\section{Experiments and Results}
\label{section:experiments-and-results}
The subsequent section delineates the experiments and the results for demonstrating the feasibility of utilizing open-source and open-access models in the aforementioned application domain. The system architecture presented in the preceding section, \Cref{section:methodologies-and-approach}, is employed to investigate and provide a comparative analysis of disparate LLMs for the purpose of exploiting user intent resolution through machine code generation.
\\\\
An implementation of the aforementioned Controller is facilitated by the Python 3 programming language. It is also employed as the base programming language for generating workflow-equivalent code due to its extensive adoption and the fact that LLMs are trained on publicly available data. Additionally, it enables the isolated and locally scoped execution of code through the \texttt{exec} function, without interfering with the global program structures of the Controller, and has the capacity to interface with the underlying execution process. This facilitates the generation of execution traces comprising function calls, their respective arguments, return values, and global context information. The generated code and the execution trace resulting from the execution of the generated code are used for evaluation. Through the integration of code blocks, the code embedded within the response of the LLM Service is decoded. The Function Table is populated with stub functions as well as real functions that implement real functionality. A complete list of the functions included in the function table is shown in \Cref{figure:functions}.
\\
\begin{figure}
    \centering
    \input{figure/functions}
    \caption{Functions accessible by the respective LLM.}
    \label{figure:functions}
\end{figure}

The Controller runs on a mobile device running the Android operating system. The \href{https://f-droid.org/en/packages/com.termux/}{Termux} and \href{https://f-droid.org/en/packages/com.termux.api/}{Termux::API} applications are used to access a shell, package manager and execution environment for running the Controller application, as well as to access to certain Android APIs via command line applications. The following open-source and open-access, as well as proprietary models are considered for the experiments:

\begin{itemize}
    \item \falconmodel
    \item \qwenmodel
    \item \phimodel
    \item \gptomodel
    \item \gptominimodel
    \item \gptturbomodel
    \item \gptpreviewmodel
\end{itemize}
Furthermore, the following user intentions, consisting of simple intentions such as smoke tests and knowledge-based as well as multi-action tasks, are taken into account: 

\begin{enumerate}[label=\arabic*., ref=\arabic*]
    \item \label{intention:1} \emph{Please sleep for 5 seconds}
    \item \label{intention:2} \emph{Please tell me a random number between 1 and 100}
    \item \label{intention:3} \emph{Please tell me the current temperature}
    \item \label{intention:4} \emph{Play a random song in my list for 5 seconds}
    \item \label{intention:5} \emph{Which is the largest city in Germany?}
    \item \label{intention:6} \emph{Please tell me all files in the current directory}
    \item \label{intention:7} \emph{Please send my car title to my insurance company}
    \item \label{intention:8} \emph{Please summarize the Wikipedia article}\\
    \emph{https://en.wikipedia.org/wiki/Transformer\_(deep\_learning\_architecture)}
    \item \label{intention:9} \emph{Please install nginx on the machine with the address 127.0.0.1:2222 running Debian GNU/Linux}
\end{enumerate}

\begin{figure}
    \centering
    \begin{minipage}{0.3\textwidth}
        \centering
        \includegraphics[width=\textwidth]{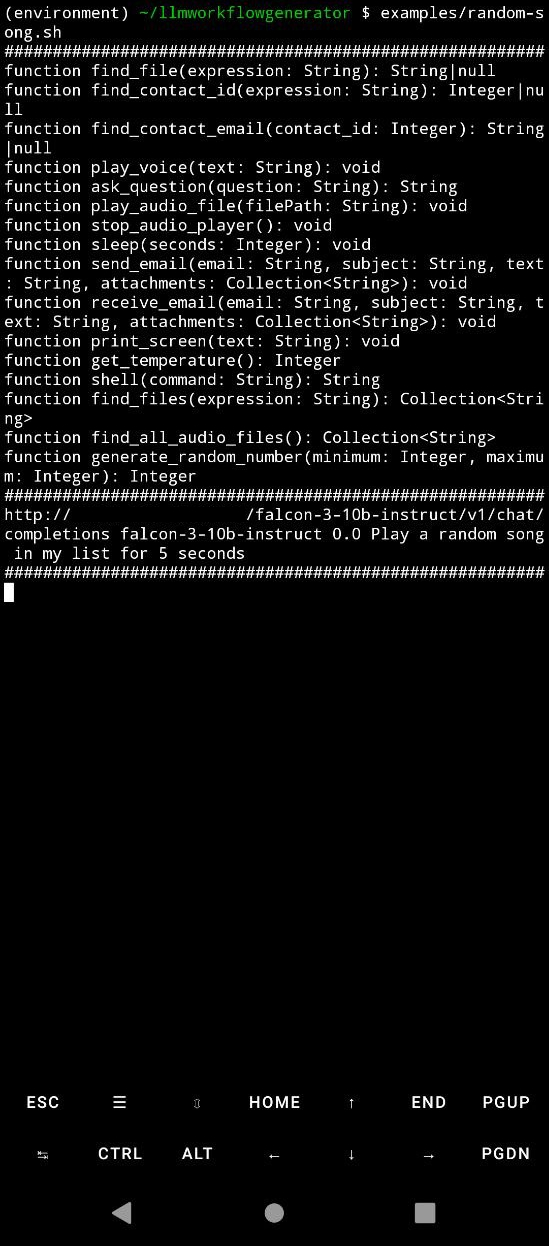}
    \end{minipage}%
    \hfill
    \begin{minipage}{0.3\textwidth}
        \centering
        \includegraphics[width=\textwidth]{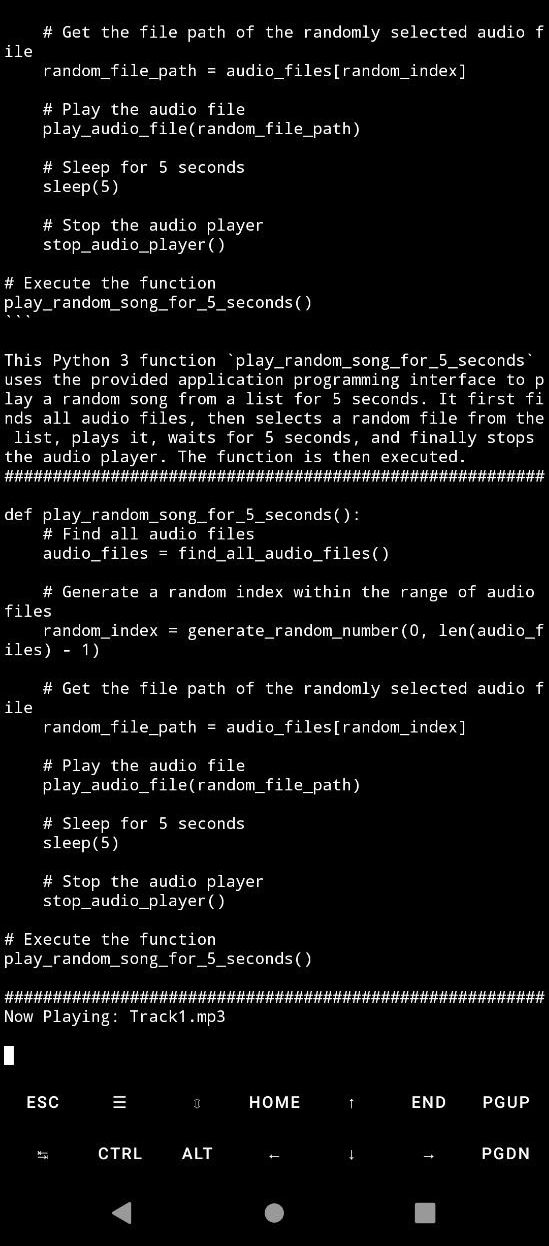}
    \end{minipage}%
    \hfill
    \begin{minipage}{0.3\textwidth}
        \centering
        \includegraphics[width=\textwidth]{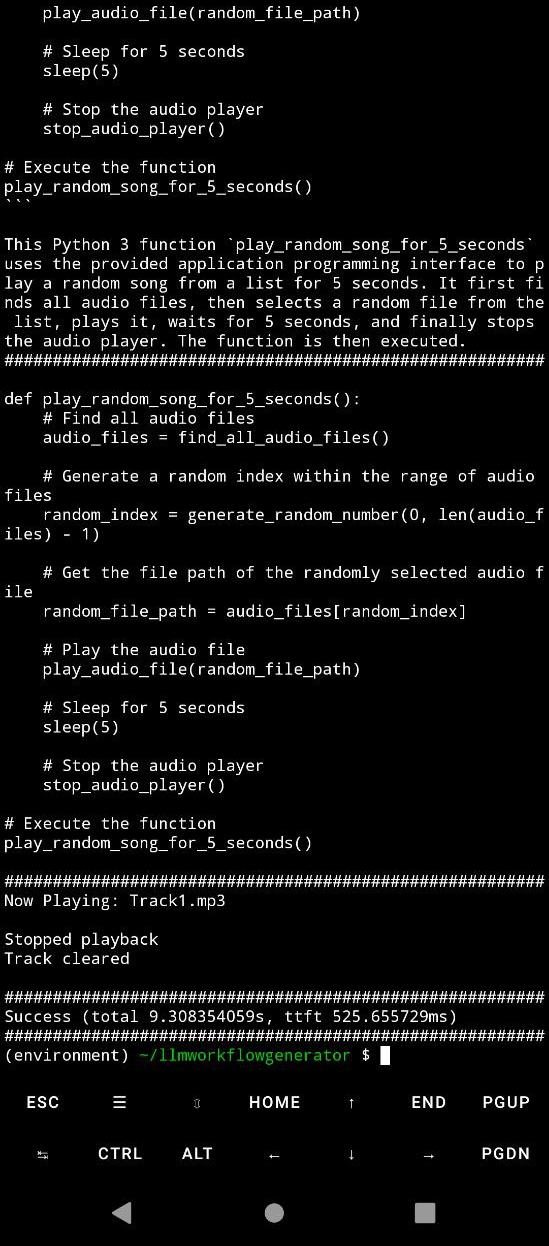}
    \end{minipage}
    \caption{Exemplary execution trace for Intention \ref{intention:4} using \falconmodel.}
    \label{figure:intention4}
\end{figure}

The selected intentions encompass a wide spectrum of capabilities and scenarios. Simple baseline functions (\ref{intention:4}, \ref{intention:2}, \ref{intention:5}) ensure that fundamental responses function correctly. External information requests (\ref{intention:3}, \ref{intention:8}) test connections to both dynamic and static knowledge sources. System-oriented tasks (\ref{intention:6}, \ref{intention:9}) simulate realistic use cases in IT and development contexts. Media as well as everyday interactions (\ref{intention:4}, \ref{intention:7}) address practical assistance functions, including security and privacy aspects. Collectively, these elements constitute a representative test set that encompasses a wide spectrum of cases, ranging from trivial to complex, security-critical, and highly practical scenarios. The Controller is configured to utilize each of the LLMs that have been presented, and is fed with each of the user intentions that have been previously outlined. The model temperature is set to $0.0$ for more deterministic results and the role to \textit{You are a Python 3 code generator} for ensuring the response consists of executable Python 3 code. The generated code and its execution traces, which are produced by the execution of the aforementioned code, are subsequently utilized for further evaluation. Each intention is transmitted to each LLM once. An exemplary resolution of Intention \ref{intention:4} employs the \falconmodel model. \Cref{figure:intention4} illustrates the invocation of the Controller and the subsequent resolution of the user intention to play a random song. Additionally, it presents the provided functions and the generated code.
\\
\begin{figure}
    \centering
    \input{figure/play-audio-file}
    \caption{Implementation of the \texttt{play\_audio\_file} function, using the command line application \texttt{termux-media-player}, provided by Termux.}
    \label{figure:play-audio-file}
\end{figure}

\Cref{table:successes-failures} provides a comprehensive overview of each model, highlighting the user intention resolutions that have been successfully addressed and those that have not met expectations. A prevailing consensus emerges from the experiments, indicating the efficacy of LLMs in facilitating automatic, machine-supported user intention resolution. This consensus extends beyond proprietary models to encompass both open-source and open-access models. The reasons for failing user intention resolutions vary and depend on the particular LLM.
\\
\begin{table}
    \footnotesize
    \centering
    \begin{tabular}{|l|c|c|c|c|c|c|c|c|c|c|c|}
    \hline
    \textbf{Model} & \textbf{\ref{intention:1}} & \textbf{\ref{intention:2}} & \textbf{\ref{intention:3}} & \textbf{\ref{intention:4}} & \textbf{\ref{intention:5}} & \textbf{\ref{intention:6}} & \textbf{\ref{intention:7}} & \textbf{\ref{intention:8}} & \textbf{\ref{intention:9}} & \textbf{\cmark} & \textbf{\xmark} \\
    \hline
    \falconmodel     & \cmark & \cmark & \cmark & \cmark & \cmark & \cmark & \xmark & \cmark & \xmark & 7 & 2 \\
    \phimodel        & \cmark & \cmark & \cmark & \cmark & \cmark & \cmark & \cmark & \xmark & \xmark & 7 & 2 \\
    \qwenmodel       & \cmark & \xmark & \cmark & \cmark & \cmark & \cmark & \cmark & \xmark & \cmark & 7 & 2 \\
    \gptomodel       & \cmark & \cmark & \cmark & \cmark & \cmark & \cmark & \cmark & \xmark & \cmark & 8 & 1 \\
    \gptominimodel   & \cmark & \cmark & \cmark & \cmark & \cmark & \cmark & \cmark & \xmark & \cmark & 8 & 1 \\
    \gptturbomodel   & \xmark & \cmark & \cmark & \cmark & \cmark & \cmark & \cmark & \xmark & \cmark & 7 & 2 \\
    \gptpreviewmodel & \cmark & \cmark & \cmark & \cmark & \cmark & \cmark & \cmark & \xmark & \cmark & 8 & 1 \\
    \hline
\end{tabular}
    \caption{Synopsis of successful (\cmark) and unsuccessful (\xmark) user intention resolutions that were achieved by the employed LLM.}
    \label{table:successes-failures}
\end{table}

The findings of the present study demonstrate that open-source and open-access models \falconmodel, \phimodel and \qwenmodel and encompass seven out of nine intention resolutions. This is on par with the proprietary model \gptturbomodel. For the other proprietary models \gptomodel, \gptominimodel and \gptpreviewmodel eight intention resolutions succeed. 
\\\\
\qwenmodel has issues with the elementary intention \ref{intention:2}, since, it utilizes the \texttt{ask\_question} function. These finding suggests that there are issues with the correct interpretation of the user intention as well as the given task. Furthermore, the elementary intention \ref{intention:1} is not fulfilled by the proprietary LLM \gptturbomodel. The code that is generated is accurate. However, the generated function is not invoked, despite being proactively instructed to do so in the provided prompt.
\\\\
\falconmodel fails with intention \ref{intention:7}, since it responds with an incorrect code block marker, using \texttt{<|assistant|>} instead of \texttt{python\textasciigrave\textasciigrave\textasciigrave} for code initiation. Notwithstanding, the resulting code is both accurate and profound. The LLM effectively addresses the user's intended purpose by employing control structures for error handling. However, intention \ref{intention:9} fails due to interpretation issues. For this intention, \falconmodel utilizes the wrong set of functions for resolving the provided user intention. It should be noted that \phimodel experiences difficulties with intention \ref{intention:9} as well. The LLM utilizes the \texttt{query\_llm} function to query itself for the required command. However, it does not successfully extract the command from the response. It is noteworthy that the command is executed directly without the response, and the \texttt{shell} function for command invocation is incorporated correctly.
\\\\
In general, the resolution of intention \ref{intention:8} is unsuccessful for all aforementioned LLMs. Although the LLM \falconmodel is successful, it employs an alternative method for achieving that success. Initially, it was hypothesized that the LLMs would employ the \texttt{http\_get\_request}  function to retrieve article content. Subsequently, the \texttt{query\_llm} function would be utilized for the purposes of reading, comprehending, and creating an article summary. This approach is not applicable in the case of the article under consideration due to its size and the inclusion of a comprehensive set of website building blocks consisting of HTML and CSS, in addition to the written text. The majority of the aforementioned LLMs respond with either a \textit{Bad Request} HTTP response or some other HTTP client error response upon invoking the \texttt{query\_llm} function, as the inputs exceed the context windows. It is determined that the \falconmodel does not employ the \texttt{http\_get\_request} function. Rather, it passes the intention directly to the \texttt{query\_llm} function for generating a response from its own internal knowledge.
\\\\
A notable finding pertains to preambles and postambles. The open-source and open-access models, namely \falconmodel, \phimodel and \qwenmodel, do not incorporate any preambles or postambles. It is demonstrated that the proprietary LLMs \gptturbomodel and \gptominimodel correctly exclude preambles and postambles as well. However, \gptpreviewmodel and \gptomodel include them. A detailed overview for each user intention resolution is shown in \Cref{table:preamble-postamble}.
\\
\begin{table}
    \footnotesize
    \centering
    \begin{tabular}{|l|c|c|c|c|c|c|c|c|c|c|c|}
    \hline
    \textbf{Model} & \textbf{\ref{intention:1}} & \textbf{\ref{intention:2}} & \textbf{\ref{intention:3}} & \textbf{\ref{intention:4}} & \textbf{\ref{intention:5}} & \textbf{\ref{intention:6}} & \textbf{\ref{intention:7}} & \textbf{\ref{intention:8}} & \textbf{\ref{intention:9}} & \textbf{\cmark} & \textbf{\xmark} \\
    \hline
    \falconmodel     & \xmark & \xmark & \xmark & \xmark & \xmark & \xmark & \xmark & \xmark & \xmark & 0 & 9 \\
    \phimodel        & \xmark & \xmark & \xmark & \xmark & \xmark & \xmark & \xmark & \xmark & \xmark & 0 & 9 \\
    \qwenmodel       & \xmark & \xmark & \xmark & \xmark & \xmark & \xmark & \xmark & \xmark & \xmark & 0 & 9 \\
    \gptomodel       & \cmark & \cmark & \cmark & \cmark & \cmark & \cmark & \cmark & \cmark & \cmark & 9 & 0 \\
    \gptominimodel   & \xmark & \xmark & \xmark & \xmark & \xmark & \xmark & \xmark & \xmark & \xmark & 0 & 9 \\
    \gptturbomodel   & \xmark & \xmark & \xmark & \xmark & \xmark & \xmark & \xmark & \xmark & \xmark & 0 & 9 \\
    \gptpreviewmodel & \cmark & \cmark & \cmark & \cmark & \cmark & \cmark & \cmark & \cmark & \cmark & 9 & 0 \\
    \hline
\end{tabular}
    \caption{Overview of preamble and postamble including (\cmark) and excluding (\xmark) user intention resolutions.}
    \label{table:preamble-postamble}
\end{table}

As illustrated in \Cref{table:code-comments}, the following provides a comprehensive overview of the user intentions for which the particular LLM includes code comments. According to the data presented, for \falconmodel, no discernible trends are identified. However, the \phimodel includes comments for each user intention resolution. \qwenmodel includes code comments a total of 3 times, showing that it tends more toward exclusion. In general, the proprietary models developed by OpenAI have a tendency to incorporate code comments. While the \gptomodel model exhibits a single instance of excluding comments, the \gptominimodel demonstrates a threefold occurrence of such exclusion. As was the case with \phimodel, \gptturbomodel, and \gptpreviewmodel incorporate them for all 9 user intention resolutions. Intention \ref{intention:8} merits particular attention, as it is noteworthy that all LLMs incorporate code comments, despite the aforementioned challenges in addressing them.
\\
\begin{table}
    \footnotesize
    \centering
    \begin{tabular}{|l|c|c|c|c|c|c|c|c|c|c|c|}
    \hline
    \textbf{Model} & \textbf{\ref{intention:1}} & \textbf{\ref{intention:2}} & \textbf{\ref{intention:3}} & \textbf{\ref{intention:4}} & \textbf{\ref{intention:5}} & \textbf{\ref{intention:6}} & \textbf{\ref{intention:7}} & \textbf{\ref{intention:8}} & \textbf{\ref{intention:9}} & \textbf{\cmark} & \textbf{\xmark} \\
    \hline
    \falconmodel     & \xmark & \xmark & \xmark & \cmark & \cmark & \cmark & \xmark & \cmark & \xmark & 4 & 5 \\
    \phimodel        & \cmark & \cmark & \cmark & \cmark & \cmark & \cmark & \cmark & \cmark & \cmark & 9 & 0 \\
    \qwenmodel       & \cmark & \xmark & \xmark & \xmark & \xmark & \xmark & \cmark & \cmark & \xmark & 3 & 6 \\
    \gptomodel       & \xmark & \cmark & \cmark & \cmark & \cmark & \cmark & \cmark & \cmark & \cmark & 8 & 1 \\
    \gptominimodel   & \xmark & \cmark & \cmark & \xmark & \cmark & \cmark & \cmark & \cmark & \xmark & 6 & 3 \\
    \gptturbomodel   & \cmark & \cmark & \cmark & \cmark & \cmark & \cmark & \cmark & \cmark & \cmark & 9 & 0 \\
    \gptpreviewmodel & \cmark & \cmark & \cmark & \cmark & \cmark & \cmark & \cmark & \cmark & \cmark & 9 & 0 \\
    \hline
\end{tabular}
    \caption{Overview of code comments including (\cmark) and excluding (\xmark) user intention resolutions.}
    \label{table:code-comments}
\end{table}

\Cref{table:average-metrics} presents the mean metrics of the Response Time and the Time to First Token. In \Cref{table:metrics-ranking} the leading amount for the particular metric is indicated. The presentation of these models is accompanied by an examination of both inclusion and exclusion, utilizing proprietary models from OpenAI. \gptomodel provides the most rapid response time. With the exception of proprietary models, the \qwenmodel generally exhibits the most rapid response time for the majority of user intentions. With respect to the response time, the \gptpreviewmodel model demonstrates the slowest performance. With the exception of the proprietary models from OpenAI, \phimodel shows the slowest performance for the majority of user intentions. For the time to first token metric, the \falconmodel offers the optimal performance, both with and without the consideration of the proprietary models, as it provides the most expeditious time to first token for each resolution. A thorough investigation into the slowest time to first token with the incorporation of the proprietary models reveals that the \gptpreviewmodel model manifests in 6 out of 9 instances, while the \gptturbomodel emerges in the remaining 3 cases. Excluding the proprietary models reveals that the \phimodel provides the slowest Time to First Token metric for the majority of 6 cases, while the \qwenmodel leads for the other 3 cases.
\\
\begin{table}
    \footnotesize
    \centering
    \begin{tabular}{|l|r|r|}
    \hline
    \textbf{Model} & \makecell{\textbf{$\approx$ Average}\\\textbf{Response Time (s)}} & \makecell{\textbf{$\approx$ Average Time}\\\textbf{to First Token (ms)}}\\
    \hline
    \falconmodel     & 6.39 & 353.4 \\
    \phimodel        & 7.16 & 398.4 \\
    \qwenmodel       & 3.42 & 390.6 \\
    \gptomodel       & 1.75 & 539.9 \\
    \gptominimodel   & 3.99 & 498.3 \\
    \gptturbomodel   & 6.53 & 883.1 \\
    \gptpreviewmodel & 7.24 & 900.1 \\
    \hline
\end{tabular}
    \caption{Average Response Time and Time to First Token for each LLM.}
    \label{table:average-metrics}
\end{table}

\begin{table}
    \footnotesize
    \centering
    \begin{tabular}{|l|l|l|}
    \hline
    & \textbf{GPTs Included} & \textbf{GPTs Excluded} \\
    \hline
    \makecell{\textbf{Fastest}\\\textbf{Response Time}} &
        \begin{tabular}[c]{@{}l@{}}9/9: \gptomodel\end{tabular} &
        \begin{tabular}[c]{@{}l@{}}2/9: \falconmodel\\ 1/9: \phimodel\\ 6/9: \qwenmodel\end{tabular} \\
    \hline
    \makecell{\textbf{Slowest}\\\textbf{Response Time}} &
    \begin{tabular}[c]{@{}l@{}}4/9: \gptpreviewmodel\\ 1/9: \falconmodel\\ 2/9: \gptturbomodel\\ 2/9: \phimodel\end{tabular} &
    \begin{tabular}[c]{@{}l@{}}2/9: \falconmodel\\ 7/9: \phimodel\end{tabular} \\
    \hline
    \makecell{\textbf{Fastest Time}\\\textbf{to First Token}} &
    9/9: \falconmodel &
    9/9: \falconmodel \\
    \hline
    \makecell{\textbf{Slowest Time}\\\textbf{to First Token}} &
    \begin{tabular}[c]{@{}l@{}}6/9: \gptpreviewmodel\\ 3/9: \gptturbomodel\end{tabular} &
    \begin{tabular}[c]{@{}l@{}}6/9: \phimodel\\ 3/9: \qwenmodel\end{tabular} \\
    \hline
\end{tabular}
    \caption{Count of leading metrics.}
    \label{table:metrics-ranking}
\end{table}

\begin{figure}
    \centering
    \includegraphics[width=0.75\linewidth]{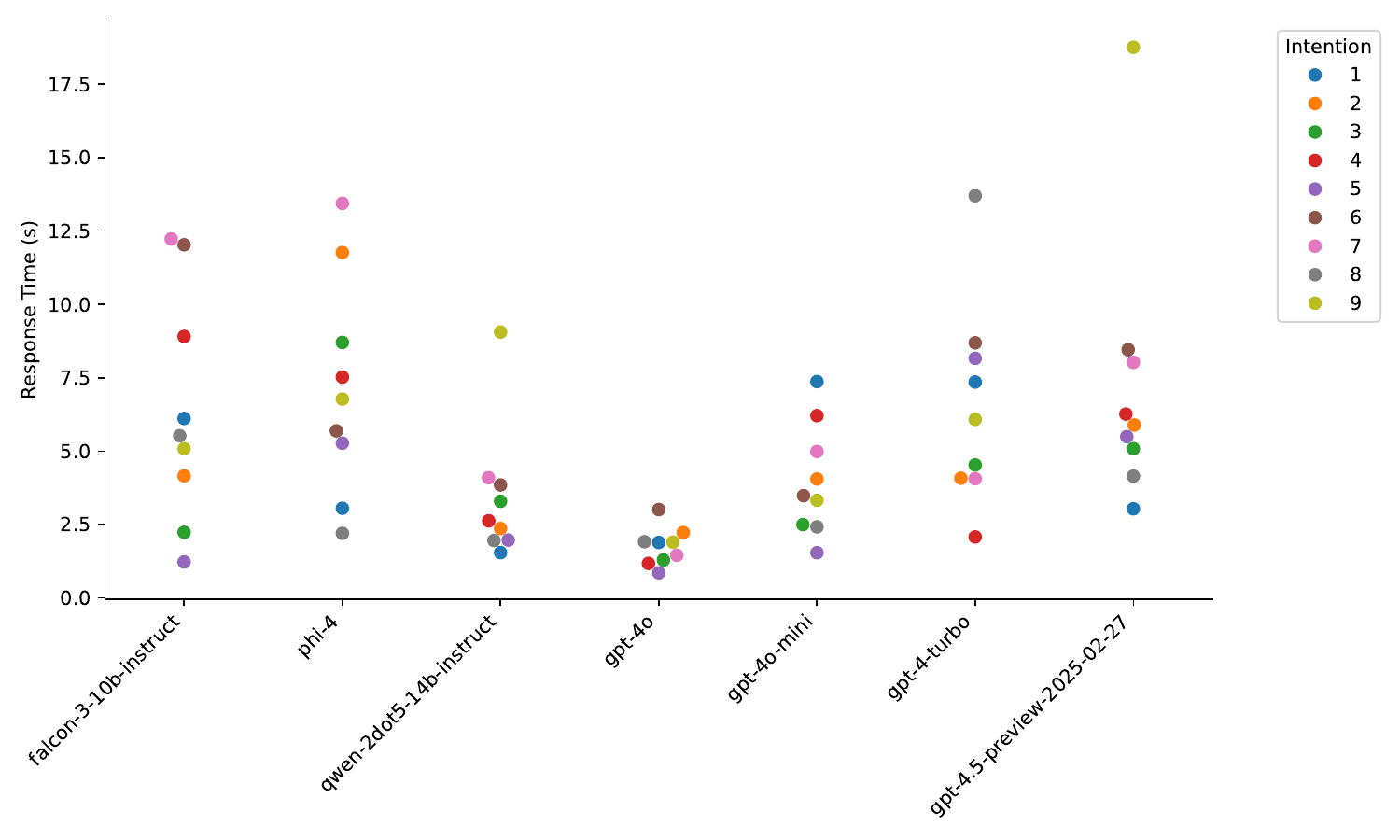}
    \caption{Response Time for each LLM and the respective user intention resolution.}
    \label{figure:response-time}
\end{figure}

\begin{figure}
    \centering
    \includegraphics[width=0.75\linewidth]{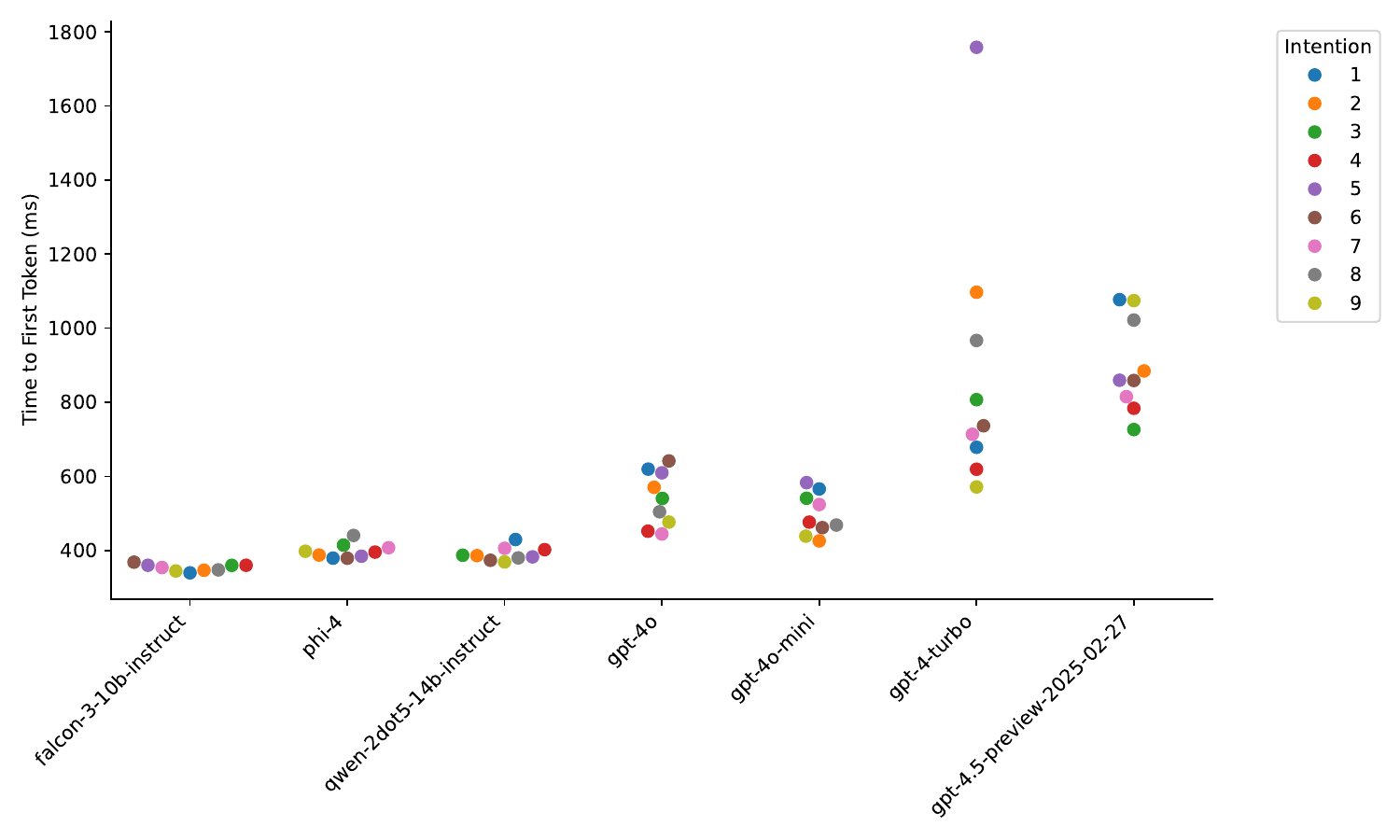}
    \caption{Time to First Token for each LLM and the respective user intention resolution.}
    \label{figure:time-to-first-token}
\end{figure}

A thorough examination of the Response Time and Time to First Token metrics, meticulously grouped by the specific LLM and the user intention, is elucidated in \Cref{figure:response-time} and \Cref{figure:time-to-first-token}. The findings indicate that the average performance of the \falconmodel model is marginally superior to that of the \phimodel model with respect to Response Time. Nonetheless, both models demonstrate deficiencies when compared with the superior \qwenmodel model, which approaches parity with the proprietary \gptomodel model. For the majority of user intention resolutions, the performance of the \falconmodel model and the \phimodel model is comparable to that of the \gptturbomodel model and the \gptpreviewmodel model. However, the performance of the \gptturbomodel model and the \gptpreviewmodel model is slightly inferior to that of the \gptominimodel model. A comparative analysis reveals that the open-access and open-source LLMs, namely \falconmodel, \phimodel, and \qwenmodel, demonstrate superior performance to proprietary models from OpenAI regarding the Time to First Token. A comparison of the performance of the models reveals that the \gptomodel and the \gptominimodel demonstrate comparable results. Conversely, the \gptturbomodel and the \gptpreviewmodel exhibit substandard performance.
\\

\section{Discussion and Result Interpretation}
\label{section:discussion-and-result-interpretation}
The experiments and results presented in \Cref{section:experiments-and-results} demonstrate that the semantic quality of responses is contingent on the specific user intention. It has been observed that none of the aforementioned LLMs demonstrate the capacity to adequately address all the user intentions provided. While proprietary models from OpenAI demonstrate a slight advantage, with an average of one additional successful outcome, the findings unmistakably underscore the substantial progress achieved by open-source and open-access models. In addressing the previously formulated research question, the present study offers a demonstration of the feasibility of employing open-access and open-source models as intermediate and middleware components for decomposing given user intentions into workflows. From a semantic perspective, the experimental models under consideration facilitate the decomposition of user intentions into actionable steps with a degree of efficacy that is nearly equivalent to that of proprietary models. Despite the fact that the proprietary flagship models demonstrated leadership in terms of the average Response Time metric throughout the course of the experiments, the performance of the open-source and open-access models remained within the acceptable range of a couple of seconds. A salient detail worthy of emphasis is that each model was exposed to experimentation on merely a single instance. The objective of this study was not to establish a statistical benchmark, but rather to compare the general ability of different LLMs to translate everyday intentions into executable workflows. A single-run configuration is indicative of realistic usage patterns, wherein users typically articulate an intention on a single occasion and anticipate a response. This configuration also circumvents the potential for bias from repeated sampling, which might favor certain models.
\\\\
Subsequent endeavors in this specific application domain pertain to the optimization of the aforementioned models for the purpose of further reducing the introduced system architecture. The efficacy of LLMs is contingent upon their incorporation into local devices. However, the substantial computational demands of the inference process currently necessitate execution on remote infrastructure. The processes of pruning, distillation, and quantization offer significant opportunities for the operation of LLMs at local scale. By decreasing the model size and computational demands without a substantial compromise in performance, these techniques enable the implementation of sophisticated AI models on devices with limited resources, not exclusive to mobile devices. Collectively, these technologies facilitate enhanced accessibility, reduced operational expenditures, augmented privacy measures, and expedited response times, unveiling novel prospects for real-world, on-device AI application and enabling the focus on operating system-oriented optimization of intent-based user interaction mechanisms. In this context, open-source and open-access models assume a particularly salient role. It is imperative to acknowledge that the reduction and optimization of the aforementioned models is not the sole pivotal step. While the employment of imperative programming languages as intermediate representations for workflows functions effectively in conjunction with LLMs, the necessity arises for an all-encompassing API to address the diverse user intents. This issue must be given due consideration for future research endeavors. Furthermore, the transition of authority and decision-making capacity to LLMs and AI in general gives rise to a substantial security concern, thereby prompting the exploration of critical research domains. The deliberate or inadvertent application of LLMs has the potential to result in adverse consequences. To illustrate this point, the direct execution of generated code in the system architecture under consideration introduces a security vulnerability. It is necessary to implement significant isolation and sandboxing mechanisms, as well as utilize operating system-provided capabilities. Another exemplary vector of attack in the experiments presented is targeted around the \textit{shell} command, since it can be misused for direct access to the system, depending on the particular system configuration and setup. These findings align with the recent studies addressing the \textit{Shutdown Problem} \cite{thornley2024shutdown}\cite{thornley2024shutdown2}, which demonstrate that AI proactively undertakes measures to circumvent system shutdown. \textit{Alignment Faking} \cite{greenblatt2024alignment} further illustrates how contemporary LLMs exhibit resistance to human intervention and correction. It is imperative to devise countermeasures and techniques to circumvent potential damage that may be engendered by the integration of LLMs and AI.
\section{Related Work}
\label{section:related-work}
In the seminal paper, the \textit{Transformer} is introduced \cite{vaswani2017attention}, which is a novel deep learning model founded on self-attention, which supersedes earlier Recurrent Neural Networks (RNNs) and Convolutional Neural Networks (CNNs). The Transformer model is distinguished by its parallelisability, accelerated training, and enhanced quality in Natural Language Processing \cite{kalla2023study}\cite{minaee2024large}. This fundamental principle underlies the construction of all contemporary LLMs, including GPT, BERT, Falcon, Phi, and Qwen. A thorough examination of transformer design optimizations through the initial months of 2024 is elucidated in \cite{10.1145/3530811}. The overview encompasses \textit{FlashAttention-2}, \textit{Mixture of Experts} and \textit{Long Context Transformers}. The widespread availability of ChatGPT \cite{wubriefgpt} has led to a substantial increase in the number of applications under consideration. For instance, the application domains of public health and medicine have been the focus of study \cite{biswas2023role}\cite{thirunavukarasu2023large}\cite{nazi2024large}, as well as those of education and pedagogy \cite{firat2023chat}\cite{kasneci2023chatgpt}. The general applicability of AI necessitates its categorization, a subject that is addressed in \cite{morrisposition}. Existing solutions such as Siri, Cortana as well as Alexa influenced the application domain of personal assistants. An overview of requirements of voice user interfaces, in particular for for blind and visually impaired users, is addressed by \cite{oumard2022pardon}. The integration of LLMs for machine-oriented user intention resolution is examined in \cite{shi2024commands} and \cite{mei2024aios}, which also present \textit{AIOS}, an operating system for LLM-based agents. \cite{verhicle2024} delineates a vision for AIOS as the core of future vehicle systems research. Recent research in this particular application domain includes the training of AI to directly operate existing GUI applications \cite{liu2024autoglmautonomousfoundationagents}. Subsequent research in the context of LLMs entails the investigation of the potential of LLMs to facilitate the recognition of user intentions within dialog systems \cite{arora2024intentdetectionagellms}. A prototype tool that automatically generates business and scientific workflows using LLMs is presented in \cite{xu2024llm4workflow}. Another application of AI, particularly LLMs, is the generation of code, a subject that has been extensively studied. Tools such as GitHub Copilot provide assistance to engineers in routine tasks \cite{MORADIDAKHEL2023111734}\cite{wermelinger2023}. An evaluation of problem-solving through code generation with GPT language models has been conducted in \cite{chen2021evaluating} \cite{lin2024soen101codegenerationemulating}. Common issues associated with the utilization of LLMs, including hallucinations and erroneous code generation, are addressed by techniques that are centered around fuzzing as well as static analysis \cite{Ouyang_2024} A number of novel approaches have been developed that utilize \textit{Grammar Augmentation} \cite{ugare2024syncodellmgenerationgrammar} and the redesign of fundamental transformer decoding algorithms \cite{zhang2023planninglargelanguagemodels}. The utilization of AI and LLMs entails specific risks \cite{bender2021dangers} and necessitates a systematic taxonomy of these risks, as outlined in \cite{weidinger2022taxonomy}. Among the most critical aspects are privacy-related concerns associated with training data \cite{carlini2021extracting}, as well as the handling of sensitive user information from communication platforms. A comprehensive survey on approaches to data privacy protection is provided in \cite{yan2024protecting}. Further issues are related to linguistic biases \cite{fleisig2024linguistic}.
\section{Conclusion}
\label{section:conclusion}
This work presents a comparative analysis of various LLMs for machine-assisted resolution of user intentions. The efficacy of the open-access and open-source models \falconmodel, \phimodel, and \qwenmodel is demonstrated to be comparable to that of the proprietary fourth-generation GPT models from OpenAI, particularly in the aforementioned application domain. The experimental results indicate that while the current flagship model \gptomodel shows the shortest average response time, the collected metrics of the open-access and open-source models remained within an acceptable range. The mentioned models, namely \falconmodel, \phimodel, and \qwenmodel, are comparable to other proprietary models, such as \gptomodel, \gptturbomodel, \gptpreviewmodel. This provides a promising foundation for the future development of systems that employ self-hosted models and integrate them with LLMs to achieve greater autonomy, facilitating the translation of user intentions into workflows and their subsequent resolution. 

\section*{Declaration on Generative AI}
During the preparation of this work, the author(s) used DeepL Write in order to: Grammar and spelling check. After using these tool(s)/service(s), the author(s) reviewed and edited the content as needed and take(s) full responsibility for the publication’s content.
\bibliography{bibliography}

@article{mei2024aios,
  title={AIOS: LLM agent operating system},
  author={Mei, Kai and Li, Zelong and Xu, Shuyuan and Ye, Ruosong and Ge, Yingqiang and Zhang, Yongfeng},
  journal={arXiv e-prints, pp. arXiv--2403},
  year={2024},
  url={10.48550/arXiv.2403.16971}
}

@article{verhicle2024,
author = {Ge, Jingwei and Chang, Cheng and Zhang, Jiawei and Li, Lingxi and Na, Xiaoxiang and Lin, Yilun and Li, Li},
year = {2024},
month = {04},
pages = {1-5},
title = {LLM-Based Operating Systems for Automated Vehicles: A New Perspective},
volume = {PP},
journal = {IEEE Transactions on Intelligent Vehicles},
doi = {10.1109/TIV.2024.3399813}
}

@inproceedings{xu2024llm4workflow,
  title={Llm4workflow: An llm-based automated workflow model generation tool},
  author={Xu, Jia and Du, Weilin and Liu, Xiao and Li, Xuejun},
  booktitle={Proceedings of the 39th IEEE/ACM International Conference on Automated Software Engineering},
  pages={2394--2398},
  year={2024}
}

@misc{arora2024intentdetectionagellms,
      title={Intent Detection in the Age of LLMs}, 
      author={Gaurav Arora and Shreya Jain and Srujana Merugu},
      year={2024},
      eprint={2410.01627},
      archivePrefix={arXiv},
      primaryClass={cs.CL},
      url={https://arxiv.org/abs/2410.01627}, 
}

@article{vaswani2017attention,
  title={Attention is all you need},
  author={Vaswani, Ashish and Shazeer, Noam and Parmar, Niki and Uszkoreit, Jakob and Jones, Llion and Gomez, Aidan N and Kaiser, {\L}ukasz and Polosukhin, Illia},
  journal={Advances in neural information processing systems},
  volume={30},
  year={2017}
}

@article{10.1145/3530811,
author = {Tay, Yi and Dehghani, Mostafa and Bahri, Dara and Metzler, Donald},
title = {Efficient Transformers: A Survey},
year = {2022},
issue_date = {June 2023},
publisher = {Association for Computing Machinery},
address = {New York, NY, USA},
volume = {55},
number = {6},
issn = {0360-0300},
url = {https://doi.org/10.1145/3530811},
doi = {10.1145/3530811},
journal = {ACM Comput. Surv.},
month = dec,
articleno = {109},
numpages = {28}
}

@misc{flerlage2025machinegeneratedcoderesolutionuser,
      title={Towards Machine-Generated Code for the Resolution of User Intentions}, 
      author={Justus Flerlage and Ilja Behnke and Odej Kao},
      year={2025},
      eprint={2504.17531},
      archivePrefix={arXiv},
      primaryClass={cs.AI},
      url={https://arxiv.org/abs/2504.17531}, 
}

@ARTICLE{wubriefgpt,
  author={Wu, Tianyu and He, Shizhu and Liu, Jingping and Sun, Siqi and Liu, Kang and Han, Qing-Long and Tang, Yang},
  journal={IEEE/CAA Journal of Automatica Sinica}, 
  title={A Brief Overview of ChatGPT: The History, Status Quo and Potential Future Development}, 
  year={2023},
  volume={10},
  number={5},
  pages={1122-1136},
  doi={10.1109/JAS.2023.123618}
}

@article{biswas2023role,
  title={Role of chat gpt in public health},
  author={Biswas, Som S},
  journal={Annals of biomedical engineering},
  volume={51},
  number={5},
  pages={868--869},
  year={2023},
  publisher={Springer},
  doi={10.1007/s10439-023-03172-7}
}

@article{thirunavukarasu2023large,
  title={Large language models in medicine},
  author={Thirunavukarasu, Arun James and Ting, Darren Shu Jeng and Elangovan, Kabilan and Gutierrez, Laura and Tan, Ting Fang and Ting, Daniel Shu Wei},
  journal={Nature medicine},
  volume={29},
  number={8},
  pages={1930--1940},
  year={2023},
  publisher={Nature Publishing Group US New York}
}

@inproceedings{nazi2024large,
  title={Large language models in healthcare and medical domain: A review},
  author={Nazi, Zabir Al and Peng, Wei},
  booktitle={Informatics},
  volume={11},
  number={3},
  pages={57},
  year={2024},
  organization={MDPI}
}

@article{firat2023chat,
  title={How chat GPT can transform autodidactic experiences and open education?},
  author={Firat, Mehmet},
  year={2023},
  publisher={OSF Preprints},
  doi={10.31219/osf.io/9ge8m},
  journal={}
}

@article{kasneci2023chatgpt,
  title={ChatGPT for good? On opportunities and challenges of large language models for education},
  author={Kasneci, Enkelejda and Se{\ss}ler, Kathrin and K{\"u}chemann, Stefan and Bannert, Maria and Dementieva, Daryna and Fischer, Frank and Gasser, Urs and Groh, Georg and G{\"u}nnemann, Stephan and H{\"u}llermeier, Eyke and others},
  journal={Learning and individual differences},
  volume={103},
  pages={102274},
  year={2023},
  publisher={Elsevier}
}

@article{kalla2023study,
  title={Study and analysis of chat GPT and its impact on different fields of study},
  author={Kalla, Dinesh and Smith, Nathan and Samaah, Fnu and Kuraku, Sivaraju},
  journal={International journal of innovative science and research technology},
  volume={8},
  number={3},
  year={2023},
  doi={10.5281/zenodo.7767675}
}

@article{minaee2024large,
  title={Large language models: A survey},
  author={Minaee, Shervin and Mikolov, Tomas and Nikzad, Narjes and Chenaghlu, Meysam and Socher, Richard and Amatriain, Xavier and Gao, Jianfeng},
  journal={arXiv preprint arXiv:2402.06196},
  year={2024}
}

@inproceedings{morrisposition,
  title={Position: Levels of AGI for Operationalizing Progress on the Path to AGI},
  author={Morris, Meredith Ringel and Sohl-Dickstein, Jascha and Fiedel, Noah and Warkentin, Tris and Dafoe, Allan and Faust, Aleksandra and Farabet, Clement and Legg, Shane},
  booktitle={Forty-first International Conference on Machine Learning},
  year=2024,
  url={10.48550/arXiv.2311.02462}
}

@article{shi2024commands,
  title={From Commands to Prompts: LLM-based Semantic File System for AIOS},
  author={Shi, Zeru and Mei, Kai and Jin, Mingyu and Su, Yongye and Zuo, Chaoji and Hua, Wenyue and Xu, Wujiang and Ren, Yujie and Liu, Zirui and Du, Mengnan and others},
  journal={arXiv preprint arXiv:2410.11843},
  year={2024}
}

@misc{liu2024autoglmautonomousfoundationagents,
      title={AutoGLM: Autonomous Foundation Agents for GUIs}, 
      author={Xiao Liu and Bo Qin and Dongzhu Liang and Guang Dong and Hanyu Lai and Hanchen Zhang and Hanlin Zhao and Iat Long Iong and Jiadai Sun and Jiaqi Wang and Junjie Gao and Junjun Shan and Kangning Liu and Shudan Zhang and Shuntian Yao and Siyi Cheng and Wentao Yao and Wenyi Zhao and Xinghan Liu and Xinyi Liu and Xinying Chen and Xinyue Yang and Yang Yang and Yifan Xu and Yu Yang and Yujia Wang and Yulin Xu and Zehan Qi and Yuxiao Dong and Jie Tang},
      year={2024},
      eprint={2411.00820},
      archivePrefix={arXiv},
      primaryClass={cs.HC},
      url={https://arxiv.org/abs/2411.00820}, 
}

@article{chen2021evaluating,
  title={Evaluating large language models trained on code},
  author={Chen, Mark and Tworek, Jerry and Jun, Heewoo and Yuan, Qiming and Pinto, Henrique Ponde De Oliveira and Kaplan, Jared and Edwards, Harri and Burda, Yuri and Joseph, Nicholas and Brockman, Greg and others},
  journal={arXiv preprint arXiv:2107.03374},
  year={2021},
  doi={10.48550/arXiv.2107.03374}
}

@misc{lin2024soen101codegenerationemulating,
      title={SOEN-101: Code Generation by Emulating Software Process Models Using Large Language Model Agents}, 
      author={Feng Lin and Dong Jae Kim and Tse-Husn and Chen},
      year={2024},
      eprint={2403.15852},
      archivePrefix={arXiv},
      primaryClass={cs.SE},
      url={https://arxiv.org/abs/2403.15852}, 
}

@article{MORADIDAKHEL2023111734,
title = {GitHub Copilot AI pair programmer: Asset or Liability?},
journal = {Journal of Systems and Software},
volume = {203},
pages = {111734},
year = {2023},
issn = {0164-1212},
doi = {10.1016/j.jss.2023.111734},
url = {https://www.sciencedirect.com/science/article/pii/S0164121223001292},
author = {Arghavan {Moradi Dakhel} and Vahid Majdinasab and Amin Nikanjam and Foutse Khomh and Michel C. Desmarais and Zhen Ming (Jack) Jiang}
}

@inproceedings{wermelinger2023,
    author = {Wermelinger, Michel},
    title = {Using GitHub Copilot to Solve Simple Programming Problems},
    year = {2023},
    isbn = {9781450394314},
    publisher = {Association for Computing Machinery},
    address = {New York, NY, USA},
    url = {https://doi.org/10.1145/3545945.3569830},
    doi = {10.1145/3545945.3569830},
    booktitle = {Proceedings of the 54th ACM Technical Symposium on Computer Science Education V. 1},
    pages = {172–178},
    numpages = {7},
    location = {Toronto ON, Canada},
    series = {SIGCSE 2023}
}

@article{Ouyang_2024,
   title={An Empirical Study of the Non-determinism of ChatGPT in Code Generation},
   ISSN={1557-7392},
   url={http://dx.doi.org/10.1145/3697010},
   DOI={10.1145/3697010},
   journal={ACM Transactions on Software Engineering and Methodology},
   publisher={Association for Computing Machinery (ACM)},
   author={Ouyang, Shuyin and Zhang, Jie M. and Harman, Mark and Wang, Meng},
   year={2024},
   month=sep
}

@misc{ugare2024syncodellmgenerationgrammar,
      title={SynCode: LLM Generation with Grammar Augmentation}, 
      author={Shubham Ugare and Tarun Suresh and Hangoo Kang and Sasa Misailovic and Gagandeep Singh},
      year={2024},
      eprint={2403.01632},
      archivePrefix={arXiv},
      primaryClass={cs.LG},
      doi={10.48550/arXiv.2403.01632}, 
}

@misc{zhang2023planninglargelanguagemodels,
      title={Planning with Large Language Models for Code Generation}, 
      author={Shun Zhang and Zhenfang Chen and Yikang Shen and Mingyu Ding and Joshua B. Tenenbaum and Chuang Gan},
      year={2023},
      eprint={2303.05510},
      archivePrefix={arXiv},
      primaryClass={cs.LG},
      doi={10.48550/arXiv.2303.05510}, 
}

@article{thornley2024shutdown,
  title={The shutdown problem: an AI engineering puzzle for decision theorists},
  author={Thornley, Elliott},
  journal={Philosophical Studies},
  pages={1--28},
  year={2024},
  publisher={Springer}
}

@article{thornley2024shutdown2,
  title={The shutdown problem: Three theorems},
  author={Thornley, Elliott},
  journal={arXiv e-prints},
  pages={arXiv--2403},
  year={2024}
}

@article{greenblatt2024alignment,
  title={Alignment faking in large language models},
  author={Greenblatt, Ryan and Denison, Carson and Wright, Benjamin and Roger, Fabien and MacDiarmid, Monte and Marks, Sam and Treutlein, Johannes and Belonax, Tim and Chen, Jack and Duvenaud, David and others},
  journal={arXiv preprint arXiv:2412.14093},
  year={2024}
}

@article{jin2024comprehensive,
  title={A comprehensive survey on process-oriented automatic text summarization with exploration of llm-based methods},
  author={Jin, Hanlei and Zhang, Yang and Meng, Dan and Wang, Jun and Tan, Jinghua},
  journal={arXiv preprint arXiv:2403.02901},
  year={2024}
}

@article{xie2024human,
  title={A human-like reasoning framework for multi-phases planning task with large language models},
  author={Xie, Chengxing and Zou, Difan},
  journal={arXiv preprint arXiv:2405.18208},
  year={2024}
}

@article{zhang2023controllable,
  title={Controllable text-to-image generation with gpt-4},
  author={Zhang, Tianjun and Zhang, Yi and Vineet, Vibhav and Joshi, Neel and Wang, Xin},
  journal={arXiv preprint arXiv:2305.18583},
  year={2023}
}

@inproceedings{weidinger2022taxonomy,
  title={Taxonomy of risks posed by language models},
  author={Weidinger, Laura and Uesato, Jonathan and Rauh, Maribeth and Griffin, Conor and Huang, Po-Sen and Mellor, John and Glaese, Amelia and Cheng, Myra and Balle, Borja and Kasirzadeh, Atoosa and others},
  booktitle={Proceedings of the 2022 ACM conference on fairness, accountability, and transparency},
  pages={214--229},
  year={2022}
}

@inproceedings{bender2021dangers,
  title={On the dangers of stochastic parrots: Can language models be too big?},
  author={Bender, Emily M and Gebru, Timnit and McMillan-Major, Angelina and Shmitchell, Shmargaret},
  booktitle={Proceedings of the 2021 ACM conference on fairness, accountability, and transparency},
  pages={610--623},
  year={2021}
}

@inproceedings{carlini2021extracting,
  title={Extracting training data from large language models},
  author={Carlini, Nicholas and Tramer, Florian and Wallace, Eric and Jagielski, Matthew and Herbert-Voss, Ariel and Lee, Katherine and Roberts, Adam and Brown, Tom and Song, Dawn and Erlingsson, Ulfar and others},
  booktitle={30th USENIX security symposium (USENIX Security 21)},
  pages={2633--2650},
  year={2021}
}

@article{yan2024protecting,
  title={On protecting the data privacy of large language models (llms): A survey},
  author={Yan, Biwei and Li, Kun and Xu, Minghui and Dong, Yueyan and Zhang, Yue and Ren, Zhaochun and Cheng, Xiuzhen},
  journal={arXiv preprint arXiv:2403.05156},
  year={2024}
}

@book{nielsen1994usability,
  title={Usability engineering},
  author={Nielsen, Jakob},
  year={1994},
  publisher={Morgan Kaufmann}
}

@inproceedings{oumard2022pardon,
  title={Pardon? An overview of the current state and requirements of voice user interfaces for blind and visually impaired users},
  author={Oumard, Christina and Kreimeier, Julian and G{\"o}tzelmann, Timo},
  booktitle={International Conference on Computers Helping People with Special Needs},
  pages={388--398},
  year={2022},
  organization={Springer}
}

@article{fleisig2024linguistic,
  title={Linguistic bias in chatgpt: Language models reinforce dialect discrimination},
  author={Fleisig, Eve and Smith, Genevieve and Bossi, Madeline and Rustagi, Ishita and Yin, Xavier and Klein, Dan},
  journal={arXiv preprint arXiv:2406.08818},
  year={2024}
}

\appendix

\end{document}